\documentclass[conference]{IEEEtran}
\renewcommand{\baselinestretch}{0.9721}
\usepackage{cite}
\usepackage{amsmath,amssymb,amsfonts}
\usepackage{algorithmic}
\usepackage{graphicx}
\usepackage{textcomp}
\usepackage[table]{xcolor}
\usepackage{float}
\usepackage{caption}
\usepackage{subcaption}
\usepackage{empheq}
\usepackage{colortbl}
\usepackage{color,soul}
\usepackage{hhline}
\usepackage{siunitx}
\usepackage{textcomp}
\usepackage{enumitem}
\usepackage{url}
\setlist{leftmargin=2mm}

\usepackage[a4paper, total={184mm,239mm}]{geometry}
\def\BibTeX{{\rm B\kern-.05em{\sc i\kern-.025em b}\kern-.08em
    T\kern-.1667em\lower.7ex\hbox{E}\kern-.125emX}}
    
    \newcommand{\etal}{\textit{et al. }}
\usepackage{multirow}

\usepackage{float}

\newcommand\copyrighttext{%
  \footnotesize \textcopyright 2021 IEEE. Personal use of this material is permitted.
  Permission from IEEE must be obtained for all other uses, in any current or future
  media, including reprinting/republishing this material for advertising or promotional
  purposes, creating new collective works, for resale or redistribution to servers or
  lists, or reuse of any copyrighted component of this work in other works.
\newcommand\copyrightnotice{%
\begin{tikzpicture}[remember picture,overlay]
\node[anchor=south,yshift=10pt] at (current page.south) {\fbox{\parbox{\dimexpr\textwidth-\fboxsep-\fboxrule\relax}{\copyrighttext}}};
\end{tikzpicture}%
}}

\begin{document}

\title{Side-Channel Trojan Insertion -- a Practical Foundry-Side Attack via ECO\\[-1.1ex]}


\author{Tiago Perez, Malik Imran, Pablo Vaz and Samuel Pagliarini\\
Department of Computer Systems - Tallinn University of Technology, Tallinn, Estonia \\
Emails: \{tiago.perez,malik.imran,pablo.vaz,samuel.pagliarini\} @taltech.ee 
\\[-4.0ex]
}




\begin{minipage}{\textwidth}

 \textbf{\huge IEEE Copyright Notice} \\[1cm]

 \Large IEEE copyright notice \copyright~2021 IEEE. Personal use of this material is permitted. Permission from IEEE must be obtained for all other uses, in any current or future media, including reprinting/republishing this material for advertising or promotional purposes, creating new collective works, for resale or redistribution to servers or lists, or reuse of any copyrighted. \\[1cm]
 
 \textbf{Accepted to be published in: 2021 IEEE 48th International Symposium on Circuits and Systems (IEEE ISCAS 2021), May 22-28, 2021}
 
\end{minipage}
\newpage

\maketitle

\begin{abstract}

    
   Design companies often outsource their integrated circuit (IC) fabrication to third parties where ICs are susceptible to malicious acts such as the insertion of a side-channel hardware trojan horse (SCT). In this paper, we present a framework for designing and inserting an SCT based on an engineering change order (ECO) flow, which makes it the first to disclose how effortlessly a trojan can be inserted into an IC. The trojan is designed with the goal of leaking multiple bits per power signature reading. Our findings and results show that a rogue element within a foundry has, today, all means necessary for performing a foundry-side attack via ECO.

\end{abstract}

\begin{IEEEkeywords}
hardware security, manufacturing-time attack, hardware trojan horse, side-channel attack, VLSI, ASIC.
\end{IEEEkeywords}



\section{Introduction} \label{sec:introduction}

  The ever-increasing cost to build high-end semiconductor manufacturing facilities -- estimated to cost \$15-20B \cite{Cost3nm} -- has made most design companies migrate to a fabless model. In practice, design houses can market integrated circuit (IC) solutions, but fabrication is outsourced to a third party. The practice of outsourcing can potentially affect the trustworthiness of an IC as a foundry (or a rogue element within the foundry) can manipulate the design for malicious purposes \cite{Guin2014, splitchip}.
  
  Manufacturing-time attacks can tamper an otherwise trustworthy IC by inserting malicious logic or modifying specific aspects of the manufacturing process \cite{TrojanSurvey,Rostami2014}. These kinds of modifications are often referred to as hardware trojans (HTs). HTs are designed to leak confidential information, to disrupt a system's specific functionality, or even to destroy the entire system. Various types of HTs have been recently studied \cite{MolesSDT, LightTrojan, WLTrojan, HT_silicon, ParaFaultInj,A2AnaTrojan, AmpSDLCrypto, iscas_forest, iscas_detection, iscas_maire}, demonstrating the potential threat of this type of attack. Moreover, a class of HTs has emerged for assisting side-channel attacks (SCA). Lin \etal \cite{MolesSDT} were the first to propose an architecture for assisting a power SCA. This specific type of trojan is the centerpiece of our work and in the remainder of this text is referred to as a side-channel trojan (SCT).

   An IC's operating characteristics (e.g., timing, power consumption, electromagnetic radiation, etc.) can be used as a side-channel to indirectly reveal information that should be internal to the IC. For this reason, keys of crypto cores \cite{FirstDPA} are often targeted. However, to mount a successful SCA, it is necessary to acquire a large amount of data to perform correlation on. Using SCTs, on the other hand, the attack time and complexity is drastically reduced. The disadvantage of SCTs is that they require a circuit modification at fabrication time, which we later show is an \textbf{effortless exercise} for the attacker.
  
  In \cite{LightTrojan}, two lightweight SCT architectures are proposed, both with the intent to induce power consumption in order to leak a crypto key. The first architecture makes use of an adapted code-division multiple access (CDMA) scheme to distribute the leakage of bits over time. The modulated bits are forwarded to a special ``leakage circuit'' that creates a CDMA channel over the power side-channel. The second architecture, in addition to the CDMA scheme, also implements intermediate states within the AES key schedule to facilitate a differential power analysis (DPA) attack. Both architectures are implemented in a field programmable gate-array (FGPA) platform. 
  
  A silicon validated HT is presented by the authors of \cite{HT_silicon,WLTrojan}. Their demonstration is a  cryptographic IC composed of an AES core and an Ultra-WideBand transmitter that leaks the key together with the transmission of the 128 bits of ciphertext. To broaden the scope of SCTs from dedicated crypto hardware to general-purpose processors (GPPs), an interesting architecture is described in \cite{AmpSDLCrypto}, where software models of crypto standards (AES and RSA) are executed on GPPs. A number of simple micro-architectural modifications has been described to induce information leakage via faulty computations or variations in
the latency and power consumption of certain instructions.
 
  
  Despite the encouraging results reported from the SCT studies mentioned so far, no study discusses how SCTs could be inserted \textbf{from the perspective of the attacker}. In this work, we present not only an SCT design methodology, but also a novel framework for SCT insertion. We assume that a rogue agent inside the foundry is the adversary and that he/she makes use of \textbf{readily available engineering change order (ECO)} capabilities of physical design tools.

\section{Threat Model and Attacker Capabilities} \label{sec:background}

An attacker inside the foundry has the objective of inserting malicious logic in a finalized layout. Thus, he/she enjoys access to all technology and cell libraries utilized by the victim. This is particularly true for advanced nodes where only a handful of cell libraries per node exist. We assume the attacker is capable of identifying a crypto core in a layout, which is a reasonable assumption for well-known AES implementations that are often regular. We do not assume the adversary understands the entire victim's design. Instead, we assume the adversary can recognize the layout/structure of a crypto core within a larger design. Our assumptions are in line with \cite{LightTrojan,HT_silicon}. Furthermore, we also assume the adversary: 1) is versed in IC design, 2) enjoys access to modern EDA tools. With the help of the inserted logic in the form of an SCT, the attacker will then attempt to leak confidential information via a power signature. For this reason, crypto cores are often the target in this type of attack \cite{HT_silicon,ParaFaultInj} -- this is also the scenario in our work.



\begin{figure}[tb]
    \centering
    \includegraphics[width=0.60\linewidth]{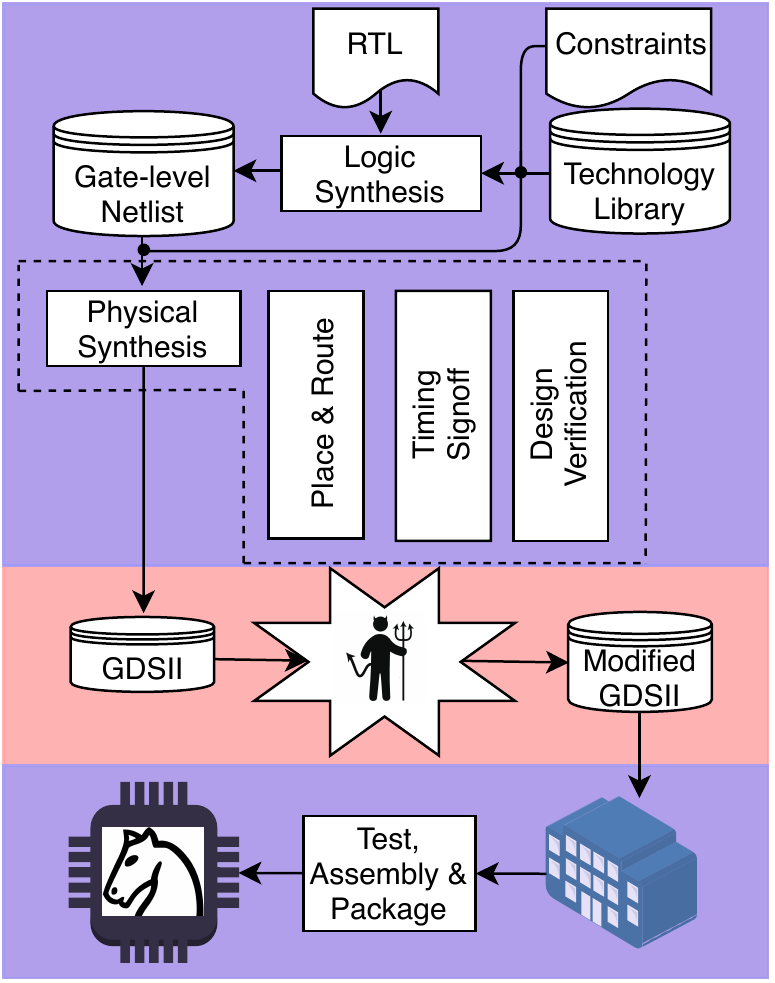}
    \caption{A typical IC design flow. Highlighted in red is the untrusted fabrication stage where the attack takes place.}
    \label{fig:phys_imp_flow}
\end{figure}

A typical physical implementation flow is described in the upper portion of Fig. \ref{fig:phys_imp_flow}. The attack takes place after the victim's layout is sent for fabrication (see red portion of Fig. \ref{fig:phys_imp_flow}). Our threat model assumes that the attacker only has access to the layout (which is the norm when outsourcing IC fabrication) -- he/she would not be able to insert the malicious logic by replicating the physical implementation flow since he/she does not have access to the RTL code, netlists, constraints, etc. 


Nevertheless, EDA tools already have the capability to deal with finalized designs. This functionality is a feature referred to as ECO. Thus, an attacker holding only the layout could use an ECO to modify or insert additional logic in a finalized layout. An ECO flow requires four inputs: technology library, cell library, gate-level netlist, and a timing constraint. The adversary already possesses the first two, but must generate/estimate the others. A gate-level netlist can be effortlessly obtained from the victim's layout through extraction \cite{VIRTUOSOOOO, Calibres, Synopsys}, while the timing constraint can be estimated to a certain degree \cite{Torrance2011, DANA, unconventionaltiming}. Our novel trojan insertion framework is shown in Fig. \ref{fig:trojan_ins}, where these two steps are considered. 

 
\section{Side-Channel Trojan Design and Insertion} \label{sec:trojan_fw}

\subsection{Side-Channel Trojan Design} \label{sec:trojan_design}
 
\begin{figure}[t!]
    \centering
    \includegraphics[width=0.98\linewidth]{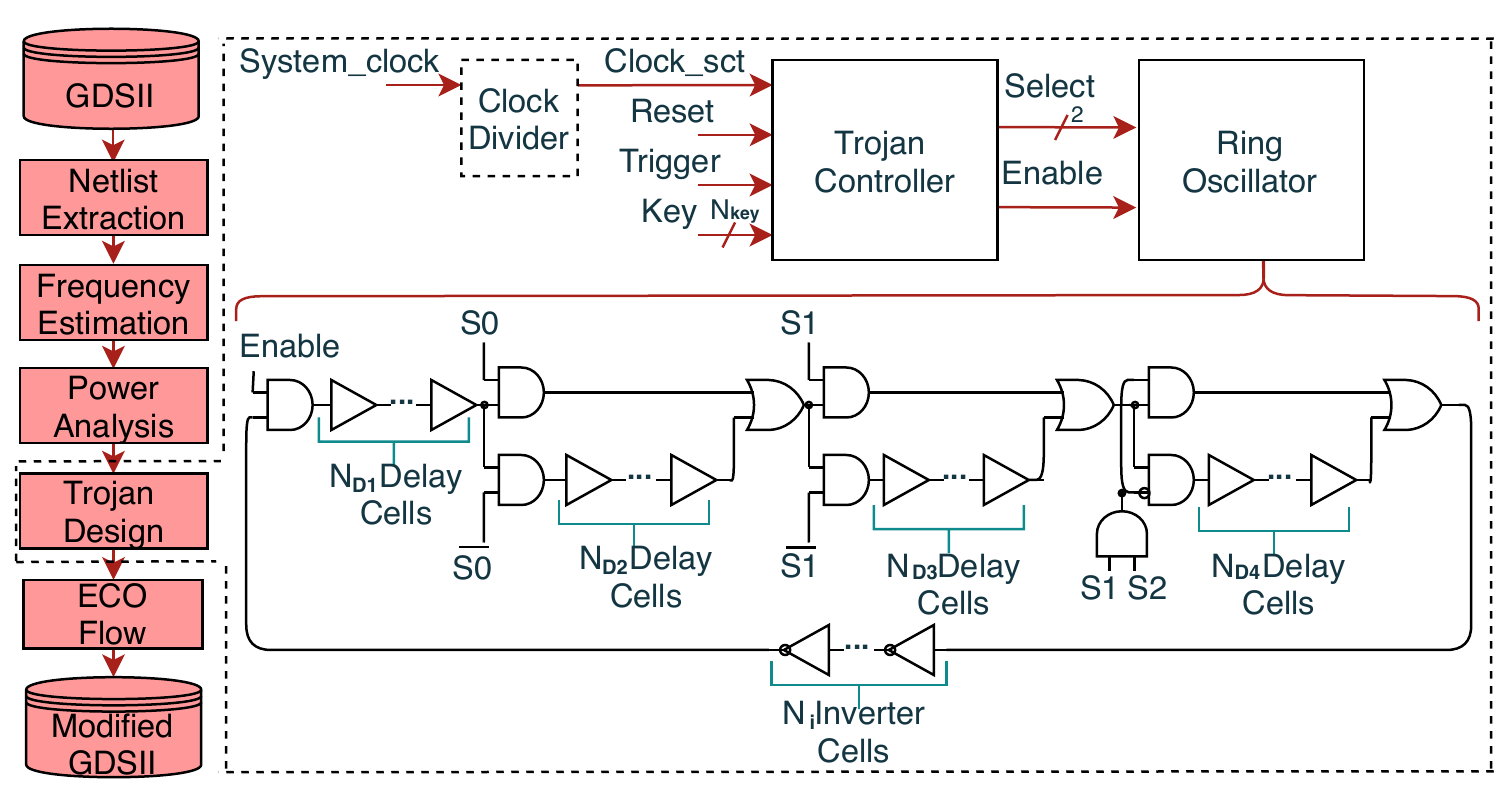}
    \caption{Our SCT insertion methodology detailed.}
    \label{fig:trojan_ins}
\end{figure}

Our proposed SCT is designed for creating an ``artificial'' yet controllable power consumption through which information is leaked. Since the majority of the power consumption in a circuit comes from the switching activity (dynamic power), a great candidate to be a power sink is a structure with a controllable frequency such as a dynamic ring-oscillator (RO). Our RO architecture implements delay stages broken into branches that are controlled by $N_{leak}$ bits. Each RO branch has two active path options: a direct connection to the next branch or a series of delay cells. The power consumption created by paths is similar to a pulse-amplitude modulation with an order equal to $2^{N_{leak}}$. An example of SCT architecture for $N_{leak}=2$ is illustrated in Fig. \ref{fig:trojan_ins}. The branch configuration is described in Table \ref{tab:rosca_op_mode}, where the leaked bits are selectors S0 and S1. 


 
 \begin{table}[htb]
        \centering
        \caption{Ring oscillator active path configuration}
        \begin{tabular}{ccccc}
        \hline 
          \textbf{S0} & \textbf{S1} & \textbf{Delay Cells} & \textbf{Inverter Cells} & \textbf{Freq.}\\
           \hline 
           0 & 0  & $N_{D1}$ & $N_i$ & High \\
           1 & 0  & $N_{D1}+N_{D2}$ & $N_i$ & Mid-high \\
           0 & 1  & $N_{D1}+N_{D3}$ & $N_i$ & Mid-low\\
           1 & 1  & $N_{D1}+N_{D2}+N_{D3}+N_{D4}$ & $N_i$ & Low \\
            \hline 
        \end{tabular}
        \label{tab:rosca_op_mode}
    \end{table}
    
A dual-sided constraint guides the attacker: he/she has to induce as much dynamic power as possible (i.e., to increase the effectiveness of the attack) while increasing as little leakage power as possible (i.e., to avoid detection). In this sense, not only the SCT has to be carefully planned, as well as when exactly will the trojan be triggered. Our approach is to not allow the trojan to compete with the dynamic power consumption of the crypto core. Therefore, when the core is actively working, the trojan is silent and the RO is not switching. When the crypto core is idle, the trojan kicks in. For this reason, our proposed SCT trojan has a \emph{Trigger} signal that is connected to the \emph{Done} signal coming from the crypto core, which marks the end of a cryptographic operation. 
 
 When triggered, the SCT connects a set of the leaking bits per clock cycle in the RO until all the $N_{key}$ bits from the crypto key are leaked. Thus, our SCT requires a connection to the system clock and reset, a trigger signal, and the crypto key. Its architecture is illustrated in Fig.~\ref{fig:trojan_ins}, consisting of three blocks: clock divider (DV), the trojan controller (TC), and the RO. The DV is required when the system clock is high and is responsible for dividing the frequency as the name suggests. Thus, the \emph{Clock\_sct} signal is either connect directly to the \emph{System\_clock} or to the DV. The TC is responsible for enabling the RO and for connecting the leaking bits in the RO. The RO starts running when the enable signal is asserted. The frequency is controlled by the select signals \emph{S0} and \emph{S1}.
 
  
 To reduce the detection probability and increase the attack's feasibility, SCTs are tailored for each target circuit. Therefore, the SCT is designed with size and power constraints, i.e., we set thresholds for the SCT based on the target's size and static power. The attacker has to acquire such information from the layout. According to Fig. \ref{fig:trojan_ins}, the layout is inspected as follows:
 
%
   
  
 \begin{itemize}
     \item[] \textbf{Netlist extraction:} since the attacker only holds the layout, a gate-level netlist has to be extracted by a CAD tool \cite{VIRTUOSOOOO, Calibres, Synopsys}.
     
     \item[] \textbf{Frequency estimation:} the attacker needs to estimate the target circuit operating frequency by performing static timing analysis on the extracted gate-level netlist. The attacker can try different clock frequencies and, by observing the critical path(s), can increase/decrease the frequency as needed until the timing slack is positive but near zero. The caveat is that multi cycle and false paths are expected to violate STA, and for this reason we say the frequency of operation is \emph{estimated}.
     
     \item[] \textbf{Power analysis:} with the extracted gate-level netlist and the estimated operating frequency, the attacker can perform a typical power analysis. For relatively large circuits, a near-accurate static power estimation can be achieved even without input vectors.
     
 \end{itemize}
  
  Therefore, after inspection, the attacker has estimated frequency and power consumption and is now ready to draw his SCT. The RO's dynamic power is tweaked by choosing an adequate number of delay cells in each individual branch as well as the number of inverter cells in the feedback path. The achieved amplitude steps have to be sufficiently different from one another for the attack to be successful.

\subsection{Side-Channel Trojan Insertion} \label{sec:trojan_insertion}

 After designing the SCT, the next step is its insertion. The attacker can utilize the ECO feature provided by commercial EDA tools for inserting the SCT. Typically, ECO is used to perform slight modifications in a finalized layout after its manufacturing (i.e., post-mask ECO). A special type of spare cell is utilized to enable ECOs. These cells do not add any functionality to the original design but, when needed, are instantiated by the ECO flow. By doing so, a new design can be generated with minimal changes in the fabrication mask set. 
 
 
 
 
For the SCT insertion via ECO, since we previously established that the attacker can discern any gate in a layout, the attacker can replace both filler and spare cells by his malicious logic \cite{Trippel2020}. Contrarily to spare cells, every layout has filler cells. During placement, EDA tools have to spread the standard cells to assure routabilility, thus mandatorily leaving gaps between cells. For more details about the relationship between placement density and HT insertion, we direct the reader to \cite{Trippel2020}.

According to Fig. \ref{fig:trojan_ins}, the ECO flow is the last step for the SCT insertion. In order to identify the filler/spare cells and remove them to create the gaps needed for the SCT, a single Cadence Innovus command is required. After the ECO, the attacker has to perform a timing sign-off to guarantee that the performance of the victim's design was kept. The SCT insertion is not likely to perturb the target's performance; it is only connected to a register (key storage) and some control signals, adding a small capacitance load. Besides, the coupling capacitance inserted by the additional routing wires is minimal due to the SCT's lightweight characteristic and the inherent goal of the ECO flow: not to disturb the existing logic. However, if the target circuit performance is perturbed, even if unlikely, it means that the size constraint used for designing the SCT was inappropriate - the adversary then proceeds to pick a different value and leak less bits per clock cycle. The attacker also has to check whether the SCT itself has timing violations. If so, the optional clock divider must be included. Every division by two requires one additional D-type flip-flop. 
    

\section{Experiments and Results} \label{sec:results}

For our experimental investigation, we have utilized AES and Present (PST) crypto cores with $N_{key}$=128 and $N_{key}$=80, respectively. AES was chosen due to its standardized status while PST was chosen due to its lightweight characteristic \cite{present_on_ASICs}. To allow the analysis of changes in \textit{frequency} and \textit{density}, the combination of these variables is explored as low-frequency low-density (LFLD), low-frequency high-density (LFHD), high-frequency low-density (HFLD), and high-frequency high-density (HFHD). Results from physical synthesis of the considered targets are presented in Table~\ref{tab:cores_phys_res}. A 65nm CMOS technology was utilized to exercise very challenging placement densities (e.g., 75\% for AES\_LFHD) and frequencies (e.g., 0.95GHz for PST\_HFLD). The values reported are for a typical corner.


 
 %

 
 \begin{table*}[htbp]
    \centering
    \caption{Physical synthesis results for our considered targets, before and after trojan insertion.}
    \begin{tabular}{p{1.3cm}p{1.2cm}|p{1.0cm}p{1cm}p{2.2cm}p{1.5cm}|p{1.0cm}p{1cm}p{2.2cm}p{1.5cm}}
    \hline 
           &    & \multicolumn{4}{c|}{\textbf{Before SCT insertion}} & \multicolumn{4}{c}{\textbf{After SCT insertion}} \\
     \textbf{Core} & \textbf{Frequency (MHz)} & \textbf{Density (\%) } & \textbf{Leakage ($\mu W$)} & \textbf{Clock Tree Power ($\mu W$)} &   \textbf{Total Power ($\mu W$)}&  \textbf{Density (\%) } &  \textbf{Leakage ($\mu W$)} & \textbf{Clock Tree Power ($\mu W$)} & \textbf{Total Power ($\mu W$)}  \\
     \hline 
        AES\_LFLD & 100    &  61  & 77.4  & 115.2  & 1670   &  63.45  & 80    & 115.8 & 1720  \\
        AES\_LFHD & 100    &  75  & 75.8  & 116.7  & 1660   &  78.20  & 79    & 117.6 & 1720  \\
        AES\_HFLD & 1000   &  58  & 1048  & 1228   & 22800  &  59.37  & 1052  & 1238  & 23015 \\
        AES\_HFHD & 1000   &  72  & 1036  & 1241   & 22610  &  73.02  & 1040  & 1252  & 22830 \\
        PST\_LFLD & 95     &  53  & 14.13 & 32.05  & 371.3 &  67.33  & 20.71 & 34.75 & 483.4  \\
        PST\_LFHD & 95     &  70  & 14.09 & 31.89  & 371.2 &  82.05  & 17.72 & 32.85 & 428.5   \\
        PST\_HFLD & 950    &  52  & 34.02 & 325.30 & 3744   &  60.89  & 36.85 & 338.1 & 4022  \\
        PST\_HFHD & 950    &  69  & 34.13 & 329.10 & 3785   &  80.26  & 36.96 & 341.5 & 4015  \\
     \hline 
    \end{tabular}
    \label{tab:cores_phys_res}
\end{table*}

 
   Based on the pre-ECO results reported in Table \ref{tab:cores_phys_res}, different SCTs were designed for each target. We assume the attacker has no means to stop clock delivery to the whole circuit, so we included the clock tree power as it has to be accounted for in our SCT power constraint. Notice how the clock tree power is significant w.r.t. the leakage power of the targets, even for the LF variants. In the results that follow, we therefore set a power budget for our SCTs of 10\% of the sum of leakage and clock tree power. Importantly, this is not a limitation of the methodology, an attacker can pick any other threshold.
   
\begin{figure}[tbp]
     \centering
     \includegraphics[width=0.6\linewidth]{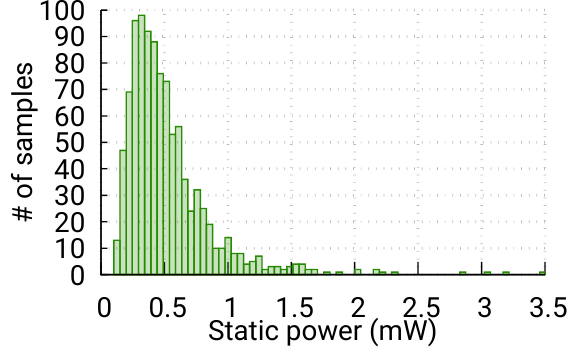}
    \caption{PST\_HFLD static power histogram, 10K MC samples.}
     \label{fig:MCsimulation}
 \end{figure}

   Aiming to obtain a better representation of the static power of the cores, we performed a Monte Carlo (MC) simulation using Cadence Spectre. This simulation was performed for 1000 samples, varying only the process with temperature fixed to 25$^{\circ}$C. The simulation results match the values reported in Table \ref{tab:cores_phys_res} for the typical corner. Fig.~\ref{fig:MCsimulation} depicts the static power distribution of PST\_HFLD. As the SCT is implemented in the very same region of the IC as the target, we can also expect the same variation in its power. 

    Once the power constraint has been established, the attacker can proceed to estimate the multiple operating frequencies of the RO (and the associated power values that effectively leak the key). Moreover, as previously alluded, we have to take into account the placed and routed version of the SCTs. For this goal, we have taken each of our SCTs and performed a custom simulation using Cadence Spectre. The oscillation frequency and power consumption of the ROs are reported in Table \ref{tab:ro_impl_res}, where each RO has been termed with a ``DXIY'' suffix. X and Y represent the number of delay and inverter cells, respectively. Notice how we do not differentiate density in the results reported in Table \ref{tab:ro_impl_res}: either the trojan fits or it does not. The SCT design is nearly agnostic to placement density.
    
    \begin{table}[tbp]
    \centering
    \caption{RO operating frequency and power consumption}
    \begin{tabular}{p{0.85cm}p{1.25cm}p{1.1cm}p{1.1cm}p{1.1cm}p{1.1cm}}
    \hline 
        \textbf{Target} & \textbf{RO} & \multicolumn{4}{c}{\textbf{Power \& Frequency ($\mu$W \& MHz)}} \\
        \multicolumn{2}{l}{\textbf{core}} & \textbf{S=00} & \textbf{S=01} & \textbf{S=10} & \textbf{S=11}  \\
    \hline
      AES\_LF     & $RO_{D6I10}$  & 19.52@65  & 16.89@45  & 14.94@34  & 12.96@20 \\
      AES\_HF     & $RO_{D10I10}$ & 198.4@551 & 182.5@483 & 160.7@390 & 139.8@300 \\
      PST\_LF & $RO_{D6I4}$   & 15.95@112 & 11.55@58  & 10.22@39  & 8.7@20 \\
      PST\_HF & $RO_{D8I10}$  & 42.02@79  & 35.56@61  & 30.88@46  & 25.66@31 \\
    \hline 
    \end{tabular}
    
    \label{tab:ro_impl_res}
\end{table}

\begin{figure}
    \centering
    \includegraphics[width=0.75\linewidth]{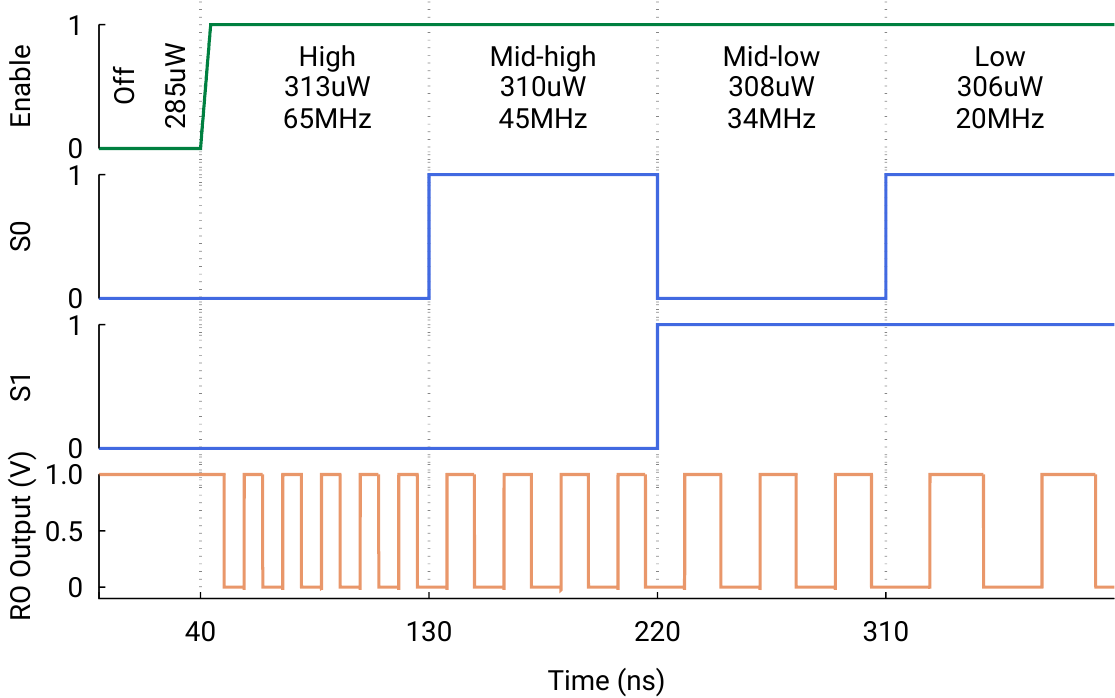}
    \caption{Side channel trojan functionality example for the AES\_LFHD, where the SCT utilizes the $RO_{D6I10}$.}
    \label{fig:ro_example}
\end{figure}

    A visual representation of how the SCT performs is given in Fig \ref{fig:ro_example}. The set of leaked keys in the image is \{00-10-01-11\} and the target circuit is AES\_LFHD. We also highlight an extreme case in the $RO_{D6I4}$ which targets the PST\_LF core. Here, the SCT alone represents about 10\% of the size of the PST core. Since area and leakage have a linear dependency, the SCT's leakage already is about 10\% of the target's leakage. Hence, the power constraint is violated. This extreme example assumes the entire IC consists of a single PST core. For a large system-on-chip containing multiple cores, the power budget for designing the SCT would be much more forgiving.
    
\begin{figure}[hb!]
       \centering
       \includegraphics[width=0.9\linewidth]{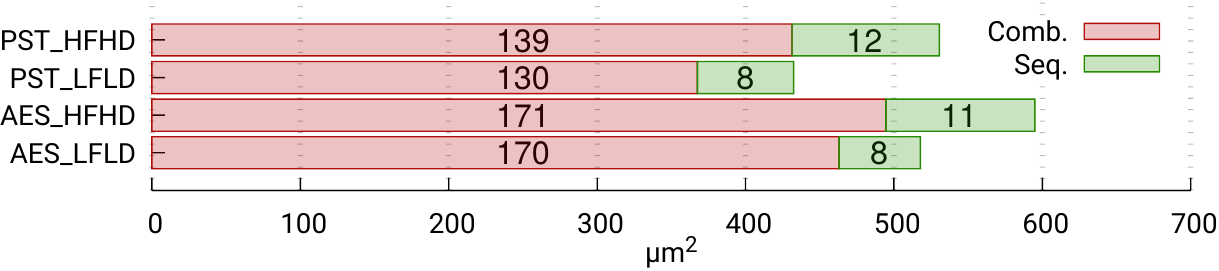}
       \caption{Comparison of area and number of cells between SCTs.}
       \label{fig:phys_trojan}
   \end{figure}

    Alongside the custom-simulated ROs, the SCTs are synthesized for each $N_{key}$ and at the same clock frequency of the target. Exclusively for the HF targets, we added the CD block to ensure the SCT does not violate timing. For AES\_HF, the system clock was divided by 8 while for PST\_HF it was divided by 16. Area and cell count for the SCTs are given in Fig. \ref{fig:phys_trojan}.

   
\begin{figure}[t]
    \centering
    \includegraphics[width=1.0\linewidth]{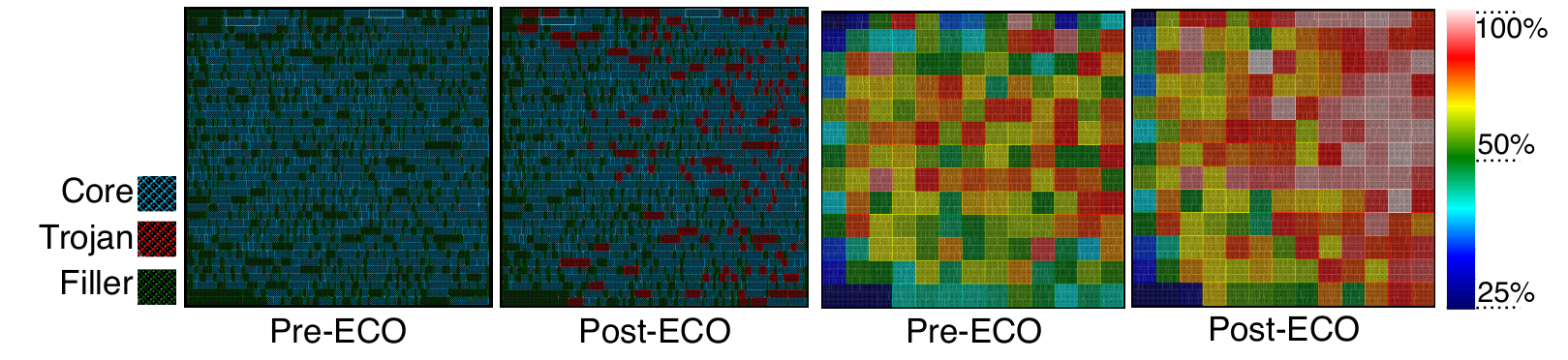}
    \caption{Placement view (left) and density map (right) of the PST\_HFHD core, before and after SCT insertion via ECO.}
    \label{fig:pst_den_place_comp}
\end{figure}

  After designing the RO and synthesizing the remainder of the SCT logic, the attacker is ready to perform the insertion via ECO. Insertion results are described on Table \ref{tab:cores_phys_res} (`After SCT insertion'). For all considered scenarios, the ECO flow was capable of placing and routing the SCT successfully, even for dense layouts. Considering that high density implies less routing resources, we verified that the ECO flow purposefully utilizes the least congested metal layers. We also provide a visual comparison of the density increase for the PST\_HFHD SCT in the left side of Fig. \ref{fig:pst_den_place_comp}. Note that the placement of the target was kept identical and only filler cells were removed during ECO. This is the key finding of this paper: \textbf{an adversary can effortlessly insert an SCT into a finalized layout}.
  
  
  

\begin{figure}[]
    \centering
    \includegraphics[width=0.95\linewidth]{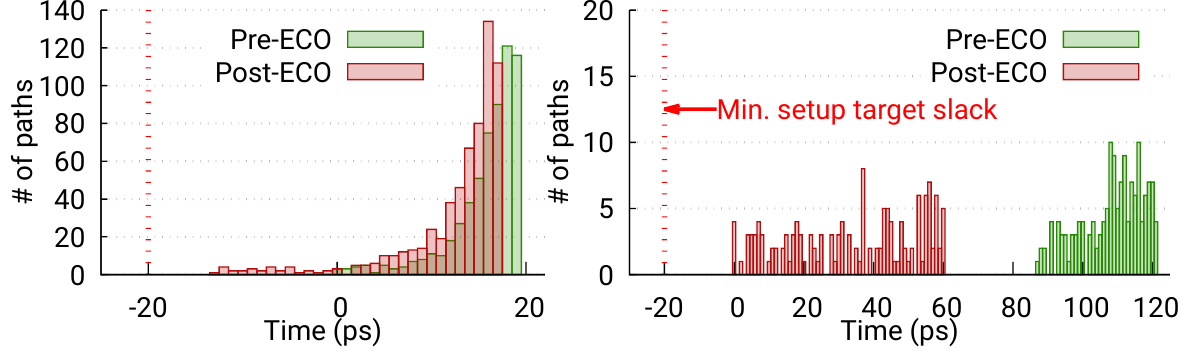}
    \caption{Pre- and post-ECO setup timing slack comparison of AES\_HFHD (right) and PST\_HFHD (left).}
    \label{fig:timing_eco}
\end{figure}

  Besides enabling the SCT insertion, the ECO flow also has to preserve the performance of the target circuit. The impact on the performance of AES\_HFHD and PST\_HFHD cores is illustrated in Fig. \ref{fig:timing_eco}. The difference in pre- and post-ECO timing slack is attributed to additional load and coupling capacitances. One can appreciate how the red bars in Fig. \ref{fig:timing_eco} are shifted to the left (w.r.t. the green bars). However, this shift was not sufficient to degrade the performance of any core. The PST\_HFHD implementation is affected slightly (which is explained by the increase in density reported in Table \ref{tab:cores_phys_res}) but does not violate our safety margin of 20ps applied to all paths. Furthermore, we argue that our proposed methodology is not only capable of inserting an SCT in a high density layout, but also of keeping the target's performance regardless of its (challenging) frequency. Finally, there are very few techniques that would assuredly counter the ECO-enabled trojan insertion \cite{splitchip, splitfab}.
  
\section{Conclusions} \label{sec:conclusion}

In this work, we proposed an SCT design methodology as well as a novel framework for SCT insertion via ECO. The SCT insertion was detailed step by step, showing that a rogue element inside a foundry can replicate it effortlessly. Furthermore, our results show how efficient an otherwise benign ECO flow can be when used for malicious reasons. Our future work includes a silicon demonstration of the inserted HT. A tapeout was completed during the writing of this paper and the fabricated ICs are expected to arrive by Jan/2021.


\newpage

\section*{Acknowledgment}

This work has been partially conducted in the project ``ICT programme'' which was supported by the European Union through the European Social Fund.
\bibliographystyle{ieeetr}
\bibliography{sd_trojan}

\end{document}